\documentclass[aps,prb,twocolumn,groupedaddress,showpacs]{revtex4}
\usepackage{dcolumn}
\usepackage{graphicx,amssymb,amsmath}
\usepackage{bm}
\usepackage{natbib}
\usepackage{epstopdf}

\usepackage{amsmath}
\usepackage{amsfonts}
\usepackage{amssymb}

\begin{document}

\title{Spin-charge separation in a strongly correlated spin-polarized chain}

\author{Shimul Akhanjee}
\author{Yaroslav Tserkovnyak}
\affiliation{Department of Physics and Astronomy, University of California, Los Angeles, California 90095, USA}

\date{\today}

\begin{abstract}
We combine the first-quantized path-integral formalism and bosonization to develop a phenomenological theory for spin-charge coupled dynamics in one-dimensional (1D) ferromagnetic systems with strong interparticle repulsion, at low temperatures. We assume an effective spin-charge separation and retain the standard Luttinger-liquid plasmon branch, which is explicitly coupled to a ferromagnetic spin-wave texture with a quadratic dispersion. The dynamic spin structure severely suppresses the plasmon peak in the single-particle propagator, in both fermionic and bosonic systems. Our analysis provides an effective theory for the new universality class of 1D ferromagnetic systems, capturing both the trapped spin and propagating spin-wave regimes of the long-time behavior.
\end{abstract}


% 71.10.Pm	Fermions in reduced dimensions (anyons, composite fermions, Luttinger liquid, etc.)
% 75.30.Ds	Spin waves
% 75.10.Pq	Spin chain models
% 71.27.+a	Strongly correlated electron systems; heavy fermions

\maketitle

\emph{Introduction.}
The Tomanaga-Luttinger liquid (TLL) is one of the most established examples of a non-Fermi liquid in low dimensions. The TLL theoretical framework captures many essential features of a large class of correlated one-dimensional (1D) quantum systems,\cite{quantum} both fermionic\cite{haldaneJPC81} and bosonic.\cite{haldane} Recent theoretical investigations have also focused on interacting 1D systems that lie outside the conventional TLL paradigm.\cite{zvonprl, zvonlong,balents,matveevPRL04,pustilnikPRL06,matveevPRL07,khodasPRL07,matveevCM07,fieteRMP07} At lower densities when the Coulomb repulsion dominates, the system becomes reminiscent of a fluctuating Wigner solid.\cite{matveevPRL04} A regime that has recently attracted a considerable interest is when the spin exchange coupling $J$ is exponentially suppressed by strong repulsions, $J \ll T\ll E_F$, where $E_F$ is the Fermi energy.\cite{zvonprl, balents,fieteRMP07} Explorations into this \textit{spin-incoherent} limit began with the pioneering Bethe ansatz calculation of Cheianov and Zvonarev,\cite{zvonprl,zvonlong} who showed that the single-particle Green's function contains exponential decay in the spin sector instead of the usual TLL power-law singularities in both spin and charge sectors. This makes this regime even more universal than the standard TLL, since the spin sector is fully thermalized. Moreover, the power-law exponents admit a nonunitary conformal and non-TLL classification.\cite{zvonprl,balents} Additionally, others have also looked for non-TLL behavior in 1D systems by exploring the band curvature effects\cite{pustilnikPRL06,khodasPRL07} beyond the customary linearized dispersions, as well as finite-energy effects stemming from an exact treatment of the spin sector, not relying on the bosonization of spin modes.\cite{matveevPRL07,matveevCM07} Very recently,\cite{zvonarevCM07} Zvonarev \textit{et al.} explored low-energy properties of strongly-correlated spin-polarized Bose gases, pointing out the existence of two regimes for spin propagation: a trapped spin regime, when spin flips are unlikely and spin excitation is mainly carried by charge fluctuations, and the open (propagating) spin-wave regime, where the spin excitations are carried by ferromagnetic magnons. Here, we develop an effective theory which helps to clarify the underlying physics and to generalize the results of Ref.~\onlinecite{zvonarevCM07}.

In this paper, we continue in the spirit of the earlier investigations and calculate the single-particle Green's function $G(x,t)$ of a 1D spin-polarized chain of either bosons or fermions, focusing on the quadratic spin-wave curvature effects. We compare the different asymptotic forms of $G(x,t)$ for both finite and vanishing exchange coupling $J$. The standard TLL model is antiferromagnetic, in accordance with the Lieb-Mattis theorem,\cite{liebmattis} which forbids spontaneous magnetization in a purely 1D quantum system. Therefore, a spin-polarized correlated system is inherently outside the scope of the conventional TLL theory. Others have discussed a possible circumvention of the Lieb-Mattis theorem in 1D systems with weak itinerant magnetism by adding a finite width to make the system effectively quasi-1D.\cite{kopietz,polar_wire} These authors have also proposed that a quasi-1D spin-polarized chain is the candidate system responsible for the $(0.5-0.7)\times 2e^2 /\hbar$ conductance anomaly. There are also strictly 1D fermionic models manifesting itinerant ferromagnetism that fall outside the class of systems considered by the Lieb-Mattis theorem.\cite{yangPRL04} One could also envision optically trapped ultracold 1D Bose gases\cite{kinoshitaNAT06} that become ferromagnetic in the presence of strong interparticle repulsion.

In our system, the band curvature effects are present in the quadratic dispersion of the ferromagnetic spin-wave excitations. We will assume a strongly interacting Wigner crystal-like limit with individual particles having very low probability of exchanging positions with their neighbors.\cite{matveevPRL04} The residual probability of tunneling through the interparticle repulsive potential gives rise to a small exchange coupling, which characterizes low-energy spin excitations. In the spin-incoherent TLL,\cite{fieteRMP07} the spin excitations are assumed to be fully thermalized by a finite temperature. Here, we will take the temperature to be below the characteristic exchange coupling, as in the usual TLL theory. The key ingredient in the following is to consider quadratic ferromagnetic rather than linear antiferromagnetic spin waves.\cite{zvonarevCM07} The spin excitations are thus treated exactly as a Heisenberg ferromagnet residing on a fluctuating lattice, with the total Hamiltonian decomposed into charge ($H_c$) and spin ($H_s$) parts, $H = H_{c}+ H_{s}$, characterized by independent, sufficiently disparate propagating velocities, $v_c$ and $v_s$, respectively. Although this
type of separation of spin and charge excitations in 1D 
Hubbard models, in the limit of very strong repulsive interactions, is supported by exact Bethe ansatz calculations,\cite{ogata} further numerical work on
more general 1D and quasi-1D systems would be desirable to establish
whether this property is more generic to strongly-coupled quasi-1D
systems. Therefore, we proceed phenomenologically, without
providing rigorous microscopic justification, noting that other authors have developed similar approaches for 1D antiferromagnetic systems.\cite{matveevPRL07,matveevCM07} 

We shall demonstrate several salient features of $G(x,t)$ in the regimes of $J>0$ and $J=0$. When $J=0$, the asymptotic form of $G(x,t)$ is characterized by strong Gaussian decay over a distance, while retaining anomalous power-law singularities with interaction-dependent exponents. Furthermore, the low-energy behavior of the tunneling density of states retains the same exponent as the spin incoherent regime,\cite{balents} $1/4g-1$, where $g$ is the TLL interaction parameter corresponding to a spinless chain with the same particle density and interactions.  This implies that the system is also characterized by non-TLL universality. For $J>0$, we find a peculiar enhancement of the plasmon peak spectral singularity that is sensitive to the spin kinematics. The effective coupling of the spin and charge collective modes is quantified in terms of a kinematic mismatch of the two coupled propagating excitations, which strongly suppresses the plasmon peak.

\emph{The effective Hamiltonian.} Let us begin with a chain of spin-$S$ particles, average interparticle spacing $a$ and strong short-range repulsion. For the entirety of this paper, we shall work within the real time and space domains, as it will facilitate the proper physical interpretation of our results. 
The charge sector retains the typical TLL form,
\begin{equation}
H_c  = \hbar v_c \int \frac{dx}{2\pi} \left[ \frac{1}{2g}{\left( {\partial _x \theta _c } \right)^2  + 2g\left( {\partial _x \phi _c } \right)^2 } \right]\,,
\end{equation}
where the bosonic $\phi_c$ and $\theta_c$ fields satisfy the usual commutation relations $[\theta _c (x),\phi _c (x^\prime)] =  - (i\pi/2){\mathop{\rm sgn}} (x - x^\prime)$, and $g$ characterizes the renormalized interparticle repulsion strength. The spin sector experiences the underlying fluctuating Wigner lattice with spins that interact via Heisenberg ferromagnetic exchange,\cite{matveevPRL04} $H_s  =- \sum_l {J_l } \vec{S}_l \cdot \vec{S}_{l + 1}$, where $l$ labels lattice sites, which are essentially distinguishable due to strong repulsion. The exchange interaction arises due to the remaining wave-function overlap between the neighboring particles.\cite{matveevPRL04} To account for the lattice zero-point motion and vibrations, in the following, the classical labels $l$ will be replaced with the bosonic displacement operators. Furthermore, to be specific, we set the nearest-neighbor exchange $J_l=\hbar J/2\pi^2S$ to be constant, where $J = \pi v_s/a$ defines a characteristic spin velocity $v_s$, and $S$ is the particle spin in units of $\hbar$. We note, however, that the exact form of the spin Hamiltonian should not be important, as long as we are dealing with a ferromagnetic ground state whose low-energy excitations are gapless magnons with quadratic dispersions. (It should also be noted that, because of the quadratic dispersion, there is no single spin velocity associated with the ferromagnetic magnons, unlike with the soundlike modes of the TLL.) Therefore, the long-wavelength free spin-wave propagator (on a stationary lattice) takes on the usual form:
\begin{equation}
G_s=\left\langle S_+(N,t)S_-(0,0)\right\rangle=S\sqrt{(2\pi/iJt)}e^{i\frac{(\pi N)^2}{2Jt}}\,.
\label{eq:spinwave}
\end{equation}

\emph{Single-particle Green's function.} In order to calculate the full Green's function, which reflects both spin and charge propagation, we have to take into account the fluctuations of our lattice. We shall make use of the first-quantized path-integral formalism to trace over the most relevant spin and charge degrees of freedom.\cite{balents} The number of sites $N(x,t)$ the spins have to propagate through is essentially the number of particles between $(0,0)$ and $(x,t)$. $N(x,t)$  fluctuates about a mean value $\bar nx$, as follows: $N(x,t) = \bar nx + \left[ {\theta _c (x,t) - \theta _c (0,0)} \right]/\pi$. The central maneuver en route to computing a particular correlation function in a given regime is to account for the explicit dependence of $N(x,t)$ on the bosonic charge degrees of freedom in the expectation brackets.

We are interested in computing the single-particle Green's function $G(x,t ) = \langle\psi _ \downarrow(x,t )\psi _\downarrow ^\dagger(0,0 )\rangle_{F_\uparrow}$, for sufficiently large $x$ and $t$, in the spin-up ferromagnetic ground state. (For spins larger than $1/2$, by ``spin down" we mean a spin $S-1$ state.) This correlator characterizes the amplitude for an injected spin-down particle to propagate from $0$ to $x$ in time $t$ in a polarized spin-up environment. In order to properly account for charge correlations, we will employ the standard bosonization dictionary. The spinless bosonic field creation and annihilation operators are given by $\psi _B^\dag  (x,t) \sim e^{ - i\phi _c (x,t)}$ and $\psi _B (x,t) \sim e^{i\phi _c (x,t)}$, where the dual field $\theta_c$ is related to the particle density fluctuations at long wavelengths by $\rho (x) = \partial _x \theta _c (x)/\pi$. It is also necessary to account for the chirality of the excitations in terms of independent right-moving and left-moving parts, which can be conveniently labeled by the light-cone variables $x_{L,R}\equiv x \pm v_c t$. We shall ignore the time ordering associated with the factor $[\theta _c (x),\phi _c (x^\prime)]$, as it simply shifts the particle number by a constant factor. 

$J=0$ \emph{bosons}. We now focus on the $J=0$ limit in a spin-up polarized ground state, where the spin-wave excitations are suppressed and only plasmons propagate, which was first discussed in Ref.~\onlinecite{zvonarevCM07}. Since the particles cannot flip their spins, the injected spin-down particle becomes effectively distinguishable and has to be removed in order to contribute to the propagator. This means that we have to force $N(x,t)$ to vanish:\cite{zvonarevCM07}
\begin{equation}
G(x,t )\sim\left\langle {\delta\left(N(x,t )\right)e^{i\left[\phi _c (x,t) - \phi _c (0,0)\right]} } \right\rangle\,.
\label{eq:j0exp}
\end{equation}
Next, we expand the $\delta$ function to obtain
\begin{equation}
G(x,t) \sim \int\limits_{ - \infty }^\infty  {\frac{{d\lambda }}{{2\pi }}} e^{ - i\lambda\frac{k_F x}{\pi} } \left\langle {e^{ - i\lambda \Theta (x,t ) + \Phi (x,t ))} } \right\rangle\,,
\end{equation}
where we have made use of the definitions $\Theta (x,t ) \equiv \theta _c (x,t) - \theta _c (0,0)$, $\Phi (x,t ) \equiv \phi _c (x,t) - \phi _c (0,0)$, and $k_F \equiv \pi\bar n$.
Furthermore, the Gaussian averages satisfy the identity $\langle {e^{ A} }\rangle  = e^{\langle {A^2 }\rangle/2} $. Therefore, after completing the squares, performing the integral over $\lambda$, and employing the standard expressions for the bosonic correlators: $\langle {\Theta ^2 } \rangle  = g\ln (x_R x_L)$, $\langle {\Phi ^2 }\rangle  = (1/4g)\ln (x_R x_L)$, $\langle {\Phi \Theta }\rangle  = (1/2)\ln(x_R/x_L)$, we finally obtain at long times
\begin{equation}
G(x,t) \sim \frac{a}{\bar{u}(t)}e^{-\frac{x^2}{2\bar{u}(t)^2}} \left( \frac{1}{x_R x_L}\right)^{\frac{1}{8g}}\,,
\label{Gdec}
\end{equation}
where $\bar{u}(t)=(a/\pi)\sqrt{2g\ln(v_ct/a)}$ is the root-mean-square particle displacement during a time $t$ due to quantum fluctuations.\cite{fietePRB05sh} One can easily notice that, because of the strong Gaussian decay on the scale of $\bar{u}$, there is no appreciable contribution to the propagator close to the light cones $x_L,x_R\approx0$ at large times (taking into account a short-distance Luttinger-liquid cutoff $\sim a$). See the $J>0$ case below, however.

$J=0$ \emph{fermions.} We must account for the Pauli exclusion principle by the negative sign acquired by the many-body wave function when two fermions are exchanged. Consequently, the factor  $\left( { - 1} \right)^N$ must be inserted in the expectation brackets (\ref{eq:j0exp}), but because the $\delta$ function enforces a vanishing $N$, for which this factor is inconsequential, we obtain the same Green's functions (\ref{Gdec}).

Clearly, the propagator (\ref{Gdec}) is dominated by a strong spatial Gaussian decay, and by inspection one can infer the asymptotic behavior of the local tunneling density of states $\nu (\epsilon)$, recovering the familiar power-law singularity near the Fermi level that is reminiscent of the spin-incoherent regime:\cite{balents} $\nu (\varepsilon ) \sim {\left| \varepsilon  \right|^{1/4g - 1} }/{\sqrt {\ln (1/\left| \varepsilon  \right|)} }$, where the energy $\varepsilon$ is measured with respect to the Fermi level.

$J > 0$ \emph{bosons ($S=1)$.} A ferromagnetic spin-wave texture is included by tracing over a product of the spin propagator (\ref{eq:spinwave}) with the standard TLL plasmon propagator. Operationally, a coupling of the two excitations is established by replacing the $\delta$ function of Eq.~(\ref{eq:j0exp}) with the spin propagator (\ref{eq:spinwave}), in terms of the bosonized particle number operator $N$:\begin{equation}
G(x,t) \sim  \frac{1}{\sqrt {iJt}} \left\langle {e^{i\frac{(\pi N)^2 }{2Jt}} \psi _B (x,t)\psi _B^\dag  (0,0)} \right\rangle\,.
\label{GJ}
\end{equation}
Evidently, the $N^2$ term gives rise to quadratic operators ${\Theta ^2 }$ in the exponent that are hard to evaluate in general. Nevertheless, as a first step in understanding the basic physics, we shall employ the leading correction to the mean-field approximation, by retaining contributions linear in the $\Theta(x,t)$ fluctuations. This approximation should be valid at large $x$ (see below for the exact condition), so that the fluctuations $\delta N\ll\bar{N}$. However, we concede that it is possible that the quadratic terms may provide meaningful corrections, and, in particular, would be required for a proper analytic continuation to imaginary time. In our approximation,
\begin{equation}
\begin{aligned}
G(x,t) &\sim \frac{1}{\sqrt {iJt}} \left\langle {e^{i \frac{{(k_F x)^2 +2k_F x\Theta (x,t)}}{{2Jt}}+i\Phi (x,t)}}  \right\rangle  \\
&\hspace{-0.5cm}= \frac{{e^{ i\frac{{(k_F x)^2 }}{{2Jt}}} }}{{\sqrt {iJt} }}\left( {\frac{{x_L}}{{x_R}}} \right)^{\frac{{k_F x}}{{2Jt}}} {\left( \frac{1}{x_R x_L } \right)^{\frac{1}{8g}+\frac{g}{2}(\frac{{k_F x}}{{Jt}})^2 } }\,,
\end{aligned}
\label{GJB}
\end{equation}
governed by the exponents
\begin{equation}
\delta_{R,L}=\frac{1}{{8g}}\pm\frac{k_Fx}{2Jt}+\frac{g}{2}\left( \frac{{k_F x}}{{Jt}}  \right)^2\,.
\label{exb}
\end{equation}

$J > 0$ \emph{fermions ($S=1/2$).} For the fermionic case, we follow similar steps, after inserting the additional factor $\left( { - 1} \right)^N= {\mathop{\rm Re}\nolimits}~{e^{i\pi N} } = (e^{i\pi N} + e^{-i\pi N})/2$ into the expectation brackets. Let us denote $G_{\pm}(x,t)$ as corresponding to the $e^{\pm i\pi N}$ parts, with $ G(x,t)= G_{+}(x,t)+ G_{-}(x,t)$:
\begin{equation}
\begin{aligned}
G_{+}(x,t) &\sim \frac{{e^{ik_F x + i\frac{{(k_F  x)^2 }}{{2 Jt}}} }}{{\sqrt { iJt} }} \\
&\times{\left( {\frac{{x_L}}{{x_R}}} \right)^{1 + \frac{{k_F x}}{{Jt}}} } {\left( \frac{1}{x_R x_L } \right)^{ \frac{1}{{8g}} +\frac{g}{2}( {1 + \frac{{k_F x}}{{Jt}}})^2  } } \,.
\end{aligned}
\label{eq:bosejf}
\end{equation}
Adding the other $G_{-}(x,t)$ component yields
\begin{equation}
G(x,t)\sim\frac{{e^{i\frac{{(k_F x)^2 }}{{2Jt}}} }}{{\sqrt {iJt} }}\left( {\frac{{e^{ik_F x}}}{{ x_R^{\delta_{R+}}x_L^{\delta_{L+}}  }} + \frac{{e^{ - ik_F x} }}{ x_R^{\delta_{R-}}x_L^{\delta_{L-}}  }} \right) \,,
\label{eq:fermjf}
\end{equation}
where
\begin{equation}
\begin{aligned}
\delta_{R\pm}=\frac{1}{{8g}}\pm1+\frac{k_Fx}{2Jt}+\frac{g}{2}\left({1\pm\frac{{k_F x}}{{Jt}} } \right)^2\,,\\
\delta_{L\pm}=\frac{1}{{8g}}\mp1-\frac{k_Fx}{2Jt}+\frac{g}{2}\left( {1\pm\frac{{k_F x}}{{Jt}} } \right)^2\,.
\end{aligned}
\label{eq:exponent}
\end{equation}
Notice that $G_-(x,t)=G_+(-x,t)$, so that $G(x,t)=G(-x,t)$, as required by the inversion symmetry.

When $J\to0$, Eq.~(\ref{GJ}) reduces to Eq.~(\ref{eq:j0exp}) and we do not have spin-wave propagation. In this regime, however, particle number $N$ is forced to vanish and our mean-field treatment assuming $\delta N\ll\bar{N}$, which is required for the validity of Eqs.~(\ref{GJB})-(\ref{eq:exponent}), fails. It is important to derive the exact conditions for which our mean-field approximation of the propagator dominates over the $N\sim0$ contribution (\ref{Gdec}). By comparing Eq.~(\ref{GJB}) with Eq.~(\ref{Gdec}), we can see that when $t\lesssim1/J=1/k_Fv_s\equiv t_\ast$, the $\propto x^2$ contribution to the exponents (\ref{exb}) actually leads to a more severe suppression of the Green's function than the Gaussian decay of Eq.~(\ref{Gdec}). [See the prefactor governed by the exponent $\delta_{L}$ of Eq.~(\ref{GP}) below, which can be rewritten as the Gaussian exponential decay $e^{-(g\ln t/2)(x/v_st)^2}$. The case of fermions is completely analogous.] Our mean-field analysis for a finite $J$ will thus fail at $t\lesssim t_\ast$, and the propagator will be dominated by the contribution (\ref{Gdec}) decaying on the scale of the interparticle spacing $a$. For $t\gg t_\ast$, Eqs.~(\ref{GJB})-(\ref{eq:exponent}) are valid, corresponding to a Gaussian decay over a longer distance $\sim v_s t$, where the spin propagation is carried predominantly by spin waves rather than charge fluctuations. Zvonarev \textit{et al.}\cite{zvonarevCM07} recently discussed the two-particle Green's function describing spin-density propagation, capturing the crossover from the trapped spin regime at $J\to0$ to propagating spin waves at a finite $J$. They found that the crossover takes place when $t\sim1/J$, irrespective of $x$, in agreement with our results.

\begin{figure}
\centerline{\includegraphics[width=\linewidth]{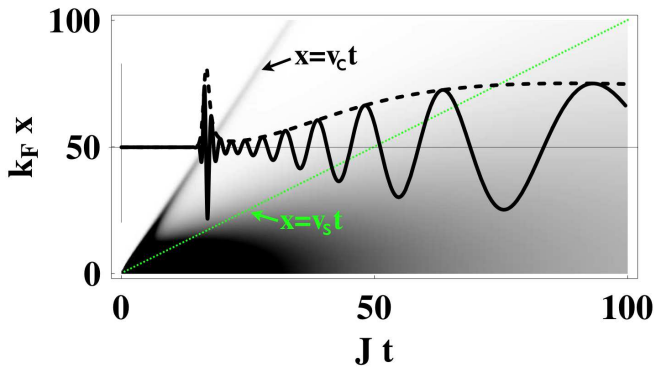}}
\caption{Single particle Green's function $G(x,t)$, Eq.~(\ref{GJB}), valid for $J t\gg1$. Here, we plot the absolute value (inserting a short-distance cutoff $a$: $x_{R,L}\to x_{R,L}\pm ia$), with white color corresponding to zero. The dashed trace is the slice of the density plot at $k_F x=50$, while the solid trace also includes the oscillations of ${\rm Re}~G(x,t)$ due to the spin propagator. Most visible features are due to the broad Gaussian peak near $x=0$ of width $\sim v_s t$ in the $x$ direction and the plasmon peak near $x=v_c t$ [governed by the exponent $\sim(g/2)(v_c/v_s)^2$, see Eq.~(\ref{GP})]. Because the relative weight of the plasmon feature is suppressed with respect to the $x=0$ correlator by the factor $\sim e^{-(g\ln t/2)(v_c/v_s)^2}$, it will only be appreciable if the ratio $v_c/v_s$ is moderate. In the plot, we chose $v_c/v_s=3$ and $g=1/2$ (corresponding to a strong short-range repulsion). Note that the exponents (\ref{exb}) approach $1/8g$ at $t\gg t_s=x/v_s$.}
\label{fig:jfinite}
\end{figure}

\emph{Discussion.} It is clear that the asymptotic forms of $G(x,t)$ discussed here do not fit within the conventional TLL form. When $J=0$, there is a strong Gaussian decay of $G(x,t)$ over relatively short distances [see Eqs.~(\ref{Gdec})],\cite{zvonarevCM07} on which is superimposed the power-law decay reminiscent of the spin-incoherent physics.\cite{balents} At a finite $J$ and $t\gg t_\ast$, the Gaussian decay is replaced by spin oscillations and some remnant plasmon features [see Eqs.~(\ref{GJB}), (\ref{eq:fermjf})].

It is important that we attempt to clarify the physical meaning of the power-law exponents (\ref{exb}) associated with the plasmon features, for a finite $J$, assuming $t\gg t_\ast$. (The case for fermions is very similar.) Let us fix the spatial coordinate $x$ and vary time $t>0$ from about $t_\ast$ though and above the plasmon peak near $x_R\sim0$ (which requires sufficiently large distances $x\gg v_ct_\ast\gg a$). The propagation of plasmon and spin-wave excitations are associated with the characteristic time scales $t_c = x/v_c$ and $t_s = x /v_s$, respectively, and for the consistency of our formulation, we require $t_s\gg t_c$. Retaining only the right-moving plasmon peak, we can approximate Eq.~(\ref{GJB}) as (omitting the obvious free spin propagator)
\begin{equation}
G\sim t^{-\delta_{L}}(t-t_c)^{-\delta_{R}}\sim e^{-\frac{g\ln t}{2}(\frac{t_s}{t})^2}(t-t_c)^{-\frac{g}{2}(\frac{t_s}{t})^2}\,,
\label{GP}
\end{equation}
assuming $t_s/t_c\gg1$ and\cite{auslaenderSCI05} $g\sim1$ (focusing on the range of $t$ from about $t_\ast$ to some multiple of $t_c$). Green's function (\ref{GP}) has two salient features: Gaussian decay in space on the scale of $v_s t$ and a remnant power-law plasmon peak at $t\to t_c$ governed by a large exponent $\propto(t_s/t_c)^2$. Let us try to understand the physical meaning of the ratio $t_s/t$ that is key to both features. First, we notice that the free spin-wave propagator, Eq.~(\ref{eq:spinwave}), enters as a prefactor in Eq.~(\ref{eq:fermjf}). When a particle is removed at the position $x$ and time $t$, the phase of the spin propagator changes by $\Delta\varphi\sim N/Jt=\pi t_s/t$. The ratio $t_s/t>1$ thus corresponds to $\Delta\varphi>\pi$, which suggests that the fluctuations of $N$ associated with the removal of a single particle lead to a destructive interference of the spin component of the full propagator, suppressing the Green's function. When $t\ll t_s$, the associated exponent scales as $(t_s/t)^2$, and when $t\sim t_\ast$, this suppression becomes more severe than that of Eq.~(\ref{Gdec}), where spins are carried by the lattice fluctuations rather than than the spin flips. This reflects an amusing interplay of the spin and charge kinematics, which has no precise analogies in the conventional TLL theory. At much longer times $t\gg t_s$, and fixed $x$, the asymptotic power-law exponent approaches $\delta_{R}+\delta_{L}\to 1/4g$ and the zero-bias spectral anomaly is governed by the exponent $1/4g-1/2$, similar to the spin-incoherent case.\cite{balents} A summary of this behavior is schematically illustrated in Fig.~{\ref{fig:jfinite}}. We finally note that the suppression of the free spin-wave correlator (\ref{eq:spinwave}) by a factor $\sim e^{-g\ln t(t_s/t)^2}$ reported in Ref.~\onlinecite{zvonarevCM07}, at $t\gg t_\ast$, appears to be also related to the ``smearing" of the oscillating spin-wave propagator by the charge fluctuations, which is responsible for the suppression of the single-particle propagator (\ref{GP}).

We are grateful to the authors of Ref.~\onlinecite{zvonarevCM07} for sharing their results prior to publication as well as for useful discussions.

\end{document}